\documentclass{emulateapj}

\begin{document}

\title{The Infrared Properties of Submillimeter Galaxies: Clues From Ultra-Deep
70$\,\mu$\lowercase{m} Imaging}


\author{Minh T. Huynh}
\affil{Spitzer Science Center, MC220-6, California Institute of Technology,
Pasadena CA 91125, USA. mhuynh@ipac.caltech.edu}

\author{Alexandra Pope}
\affil{Department of Physics and Astronomy, University of British Columbia,
Vancouver, BC, V6T 1Z1, Canada}

\author{David T. Frayer}
\affil{Spitzer Science Center, MC220-6, California Institute of Technology,
Pasadena CA 91125, USA}

\author{Douglas Scott}
\affil{Department of Physics and Astronomy, University of British Columbia,
Vancouver, BC, V6T 1Z1, Canada}

\begin{abstract}

We present 70$\,\mu$m properties of submillimeter galaxies (SMGs) in the Great
Observatories Origins Deep Survey (GOODS) North field.  Out of thirty
submillimeter galaxies ($S_{850}\,{>}\,2\,$mJy) in the central GOODS-N region,
we find two with secure 70$\,\mu$m detections.  These are the first 70$\,\mu$m
detections of SMGs.  One of the matched SMGs is at $z\,{\sim}\,0.5$ and has
$S_{70}/S_{850}$ and $S_{70}/S_{24}$ ratios consistent with a cool galaxy.  The
second SMG ($z\,{=}\,1.2$) has infrared-submm colors which indicate it is more
actively forming stars.  We examine the average 70$\,\mu$m properties of the
SMGs by performing a stacking analysis, which also allows us to estimate that
$S_{850}\,{>}\,2\,$mJy SMGs contribute $9\pm3$\% of the 70$\,\mu$m background
light.  The $S_{850}/S_{70}$ colors of the SMG
population as a whole is best fit by cool galaxies, and because of the
redshifting effects these constraints are mainly on the lower $z$ sub-sample.
We fit Spectral Energy
Distributions (SEDs) to the far-infrared data points of the two detected SMGs
and the average low redshift SMG ($z_{\rm median}\,{=}\,1.4$).  We find that
the average low-$z$ SMG has a cooler dust temperature than local ultraluminous
infrared galaxies (ULIRGs) of similar luminosity and an SED which is best fit
by scaled up versions of normal spiral galaxies.  The average low-$z$ SMG is
found to have a typical dust temperature $T\,{=}\,21$--$33\,$K and infrared
luminosity $L_{8-1000\mu m}=8.0 \times 10^{11} {\rm L}_{\odot}$.  We estimate
the AGN contribution to the total infrared luminosity of low-$z$ SMGs
is less than 23\%. 

\end{abstract}


\section{Introduction}

Deep submillimeter surveys provide a probe of galaxies which is almost
independent of luminosity for a wide redshift range  $1\,{<}\,z\,{<}\,5$,
because of the negative K-correction in the infrared (e.g.~\citealp{blain2002}).  The Submillimeter Common User
Bolometer Array (SCUBA; \citealp{holland1999}) and Max-Planck Millimeter
Bolometer (MAMBO, \citealp{kreysa1998}) have now discovered several hundred
submillimeter galaxies (SMGs: e.g.~\citealp{smail1997}, \citealp{hughes1998},
\citealp{barger1998}, \citealp{borys2002}, \citealp{chapman2002},
\citealp{webb2003}, \citealp{greve2004}, \citealp{pope2005},
\citealp{coppin2006}). 
Our knowledge of SMGs has been hampered by the large beamsize of the submm
telescopes, which makes it difficult to identify counterparts in the optical
and near-infrared.  For example, in the Great Observatories Origins Deep Survey
(GOODS) North field there are typically ${\sim}\,10$ {\sl Hubble Space
Telescope\/} ({\sl HST\/}) Advanced Camera for Surveys (ACS) optical galaxies
within a single SCUBA 15 arcsec beam. 

Deep radio surveys have proven to be one of the best ways to identify the
counterparts to SMGs (e.g.~\citealp{ivison2000}, \citealp{barger2000},
\citealp{pope2006}).  The star formation processes that heat the dust
responsible for the submm light also produce radio emission, as evidenced by
the well known far-infrared--radio correlation.
Taking advantage of this, SMGs have been matched to radio sources, which have
much better positional accuracy, and the optical/near-IR counterparts can then
be found after the radio identification has been made.  This is possible
because the density of
sources in the deepest radio images is much less than that of optical
galaxies, so the probability of a chance coincidence is lower.  

The current knowledge of SMGs is biased towards the radio identified sources.
The SMGs have a median redshift of about 2.2 and  bolometric infrared (IR)
luminosities $L_{\rm IR}\,{>}\,10^{12}\,{\rm L}_\odot$ \citep{chapman2004,
chapman2005}.  The optically identified SMGs are faint optically,
($i_{ 775}\,{\gtrsim}\,22$), and red ($i_{775}-K_{\rm s}\,{\simeq}\,2.3$ in AB
magnitudes), with about 30\% having colors consistent with Extremely Red
Objects (EROs, \citealp{pope2005}).  Thus SMGs are thought to be high redshift,
dusty analogs of local ultraluminous infrared galaxies (ULIRGs). 

The spectral energy distributions (SEDs) of SMGs have traditionally been fit
with local ULIRGs as templates, and in particular Arp220
(e.g.~\citealp{barger2000}), which has an effective dust temperature of about
42--47$\,$K (\citealp{klaas1997}, \citealp{dunne2000}).  Chapman et al.~(2005)
found a typical dust temperature of $36\pm7\,$K and a median
$L_{\rm IR}\,{=}\,8.5\times10^{12}\,{\rm L}_\odot$ for their SMG sample.  This
is cooler than local ULIRGs, which have an average dust temperature of
$43\pm6\,$K (based on the Dunne et al.~2000 sample).  The Far-IR
BACKground (FIRBACK) study of 170$\,\mu$m selected galaxies found two ULIRGs
between $0.5\,{<}\,z\,{<}\,1$ which have SEDs cooler and less luminous than
Arp220 \citep{sajina2006}.  A 350$\,\mu$m study of radio-detected SMGs found
temperatures of $35\pm3\,$K (Kovacs et al.~2006).  Using the $24\,\mu$m
imaging from the {\sl Spitzer\/} Legacy Project GOODS, \cite{pope2006}
securely identify 60\% of SMGs in this field in the mid-infrared (MIR), and
have tentative counterparts for another 34\%.  It was found that the
observed MIR--submm--radio
SED of the SMGs peak at longer wavelengths than local ULIRGs and are best fit
by models with temperatures of about $30\,$K (Pope et al.~2006).  There is thus
an emerging picture that SMGs are cooler than previously thought.

Dust temperature affects the inferred IR luminosity, and hence star formation
rate (SFR), that is derived for these galaxies.  Studies of SMGs have often
assumed dust temperatures of $40\,$K (see \citealp{blain2002}).  A drop from
$40\,$K to $35\,$K in temperature decreases the IR luminosity by about a factor
of two.  Better knowledge of the temperature of SMGs is thus crucial for more
accurate luminosity estimates. 

The advent of the {\sl Spitzer Space Telescope\/} makes it possible to study
the MIR and FIR properties of SMGs in detail for the first time.  At the median
SMG redshift of $z\,{\sim}\,2$, PAH and silicate features fall into the
rest-frame wavelength of the 24$\,\mu$m band.  This can make it difficult to
determine the total infrared luminosities and to fit model SEDs well to the
FIR, which does not always correlate with the MIR.  The 70$\,\mu$m band of the
Multiband Imaging Photometer for {\sl Spitzer\/} (MIPS) instrument is not
affected by PAH or silicate features for redshifts less than about 3, while
the 160$\,\mu$m band is not affected at all.  Moreover, these MIPS bands are
closer to the FIR peak than the MIR data points used in previous studies. 
The longer wavelength MIPS bands should therefore be extremely useful for
studying SMGs.  The sensitivities and confusion limits at 70$\,\mu$m 
make this feasible, but difficult, and hence only the deepest MIPS
data are likely to lead to SMG detections.

In this paper we present a study of the FIR properties and SEDs of submm
galaxies using the deepest 70 and 160$\,\mu$m data available for the GOODS-N
field.  In particular, we use the 70 and 160$\,\mu$m data to check if the SEDs
of distant SMGs are consistent with those of local ULIRGs. 

We assume a Hubble constant of $71\,{\rm km}\,{\rm s}^{-1}{\rm Mpc}^{-1}$,
together with matter and cosmological constant density parameters of
$\Omega_{\rm M}=0.27$ and $\Omega_{\rm \Lambda}=0.73$ in this paper. 
We also use the notation $S_{70}$, $S_{850}$, $S_{1.4}$ etc.~throughout
for the flux densities at $850\,\mu$m, $70\,\mu$m and $1.4\,$GHz.

\section{The Submillimeter Sample}

The SCUBA `super-map' of GOODS-N contains 35 robust 850$\,\mu$m detections.
Details of the data reduction and source extraction can be found in Borys et
al.~(2003) and Pope et al.~(2005).  This submm image contains all publicly
available SCUBA mapping data in this field taken up until 2004.  Although
$450\,\mu$m data were also taken, they unfortunately have essentially no
constraining power.  Of the 35
SCUBA sources, 33 have likely radio and/or {\sl Spitzer\/} counterparts using
the Very Large Array (VLA), InfraRed Array Camera (IRAC) and MIPS 24$\,\mu$m
data in GOODS-N (see Pope et al.~2006).  Thirty of these SCUBA galaxies are within
the ultra-deep 70$\,\mu$m image area.  SCUBA sources with low signal-to-noise
ratios (SNRs) may have true flux densities which are boosted by
a factor which depends on the source brightness and the local noise in the
SCUBA map, and can be estimated if the source counts are known
(see e.g.~\citealp{coppin2005}).  Therefore in this study we use the deboosted
850$\,\mu$m flux densities given in Pope et al.~(2006).

\section{Spitzer 70 and 160$\,\mu$\lowercase{m} Observations}

The MIPS 70$\,\mu$m observations were carried out during Cycle 1 of the General
Observer program ({\sl Spitzer\/} program ID 3325, Frayer et al.~2006).  The
inner $10^\prime\times10^\prime$ of GOODS-N was mapped to a depth of
$10.6\,$ksec. The data were taken in the small-field photometry mode of MIPS
with 10-second Data Collection Events (DCEs).  Each Astronomical Observation
Request (AOR) consisted of an 8-position cluster map, and the observations
were completed with 12 AORs in total.  In addition to our GO data, we used the
MIPS Guaranteed Time Observers (GTO) data (program ID 81, \citealp{dole2004}).
These GTO data have an integration time of $600\,$sec. 

The raw data were processed off-line using the Germanium Reprocessing Tools
({\tt GERT}, version S13.1), following the algorithms derived by the MIPS team
\citep{gordon2005}.  Instrumental artifacts in the Basic Calibrated Data (BCDs)
were removed using the filtering techniques adopted for the extragalactic First
Look Survey (xFLS, \citealp{frayer2006}). The data were calibrated assuming an
absolute flux calibration factor of $702\,{\rm MJy}\,{\rm sr}^{-1}$ per
MIPS-70$\mu$m unit for stellar SEDs.  The flux densities were multiplied by 1.09 to apply 
a galactic SED color correction. This assumes a power law SED of the form 
$f_\nu \propto \nu^\alpha$  and $\alpha = -1$, but the color corrections are similar for $\alpha = 0$ to $-3$. 
A more detailed discussion of the data reduction can be
found in \cite{frayer2006b}.

The final image achieves a sensitivity of ${\sim}\,0.6\,$mJy rms and we have
cataloged 101 sources (over $120\,{\rm arcmin}^2$) with $S_{70}\,{>}\,2.3\,$mJy
(SNR$\,{>}\,3\sigma$).  We catalog a region slightly larger than the area of deepest coverage 
to include some relatively bright 70 $\mu$m sources and improve statistics. 
The source counts are presented in \cite{frayer2006b}
and a full catalog will be presented in Huynh et al.~(in preparation).  The
70$\,\mu$m image has a beam size of 18\farcs5 FWHM, and in the presence of
Gaussian noise the 1$\sigma$ positional error of sources is of the order
$0.5\,\theta_{\rm FWHM}/{\rm SNR}$, i.e. 3\arcsec\  for the faintest sources. 

The 160$\,\mu$m observations of the GOODS-N region were taken as part of the
MIPS GTO program in 2004.  These data were taken in the scan mode of MIPS and we applied the standard 
filtering techniques to the 160$\,\mu$m BCDs similar to what was used for the xFLS \citep{frayer2006}.  The data were
calibrated using a factor of $44.7\,{\rm MJy}\,{\rm sr}^{-1}$ per
MIPS-160$\,\mu$m unit.   The 160$\,\mu$m data have an effective integration
time of 120 seconds and the 160$\,\mu$m image reaches a sensitivity of
${\sim}\,15\,$mJy rms. A  multiplication of 1.04 was applied to the 160$\,\mu$m flux densities to color correct for galaxy SEDs. We also note that the 160 $\mu$m light leak from 1 -- 1.6 $\mu$m is not a problem for this data because there are no blue sources with m$_{\rm J}$~$\gtrsim$~5.5 in the field.

\section{Identifications at 70 and 160$\,\mu$\lowercase{m}}

The negative K-correction at submillimeter wavelengths means that SMGs are
detectable in deep SCUBA images over a wide redshift range, $1\,{<}\,z\,{<}\,5$
(e.g.~\citealp{blain2002}).  However, at 70 and 160$\,\mu$m, the K-correction
does not compensate for the distance dimming, and the flux density of a galaxy
with a given intrinsic luminosity drops steeply as a function of redshift
(e.g.~\citealp{lagache2005}). This is reflected in the high median redshifts
of the SMGs ($z\,{\simeq}\,2.0$, Pope et al.~2006) compared to the 70$\,\mu$m
sources ($z\,{\simeq}\,0.5$, Huynh et al.~in preparation). 
Hence the SMG and 70$\,\mu$m samples are unlikely to have much overlap, and
we only expect to detect the low redshift SMGs in deep 70$\,\mu$m imaging. 

We examined the 70$\,\mu$m image for counterparts to the SCUBA sources.  To do
this the 70$\,\mu$m catalogue positions were compared to IRAC positions of the
SCUBA counterparts (Pope et al.~2006), searching within 
10\arcsec\ of each submm counterpart.  This search radius was chosen in order
to take into account the typical positional uncertainties of low SNR 70$\,\mu$m
and SCUBA sources added in quadrature.

This procedure uncovered two secure identifications of SMGs at 70$\,\mu$m,
GN26 and GN13 (following the naming convention of Pope et al.~2005).  These
sources have 70$\,\mu$m SNRs of 12 and 8, respectively, and hence 70$\,\mu$m
positional uncertainties of about 1\farcs3 and 1\farcs5 (including the
{\sl Spitzer\/} 1\arcsec\ pointing uncertainty).  The positional offset of the
70$\,\mu$m source relative to the IRAC position is 0.6 and 0.3 arcsec for GN26
and GN13, respectively, which is well within the positional uncertainty.  The
probability that one or more 70$\,\mu$m sources lies randomly within a distance
$\theta$ of a SCUBA counterpart is:
\begin{equation}
P = 1 - \exp (-\pi n \theta^2) \, , 
\end{equation}
given a surface density $n$ of 70$\,\mu$m sources (often called the
$P$-statistic, e.g.~\citealp{downes1986}).
Hence the probability of any 70$\,\mu$m source lying within 1\arcsec\ of a
SCUBA counterpart is 0.07\%, using the 70$\,\mu$m source density of 101 sources
over 120 arcmin$^2$, and hence the 70$\,\mu$m matches to GN13 and GN26 are
likely to be real.

Two more SMGs (GN12 and GN32) have a nearby 70$\,\mu$m source at distances of
4.7\arcsec and 9.5\arcsec, respectively.  The 70$\,\mu$m sources near GN12 and
GN32 have IRAC and 24$\,\mu$m counterparts which do not match the
identifications for the SCUBA source (Pope et al.~2006).  Although it is likely
that some fraction of the 70$\,\mu$m flux density is associated with the submm
galaxy, it is difficult to determine this fraction because of the other
24$\,\mu$m sources in the vicinity.  Using Equation~(1), the random probability
that 2/30 SCUBA sources have a 70$\,\mu$m source within a distance of
10\arcsec\ can be determined to be 28\%.  This is consistent with the
70$\,\mu$m sources near GN12 and GN32 being random matches. 

At 160$\,\mu$m the only SMG detected is GN26.  The beamsize at 160$\,\mu$m is
40\arcsec, so 70$\,\mu$m and/or SCUBA sources which are close together may be
blended into one 160$\,\mu$m source.  Examination of the 70$\,\mu$m MIPS image
suggests that the 160$\,\mu$m flux density of GN26 has some contribution from
another 70$\,\mu$m source (at $z\,{=}\,0.46$) not associated with GN26.  We
therefore deblended GN26 with a double Gaussian fit, fixing positions to the
IRAC counterparts of the two 70$\,\mu$m sources which contribute to the
160$\,\mu$m flux density.  We find that GN26 has $S_{160}\,{=}\,110\pm27\,$mJy,
which is 60\% of the flux density of the 160$\,\mu$m complex.  The uncertainty
at 160$\,\mu$m is conservatively estimated to be 25\%, taking into account
absolute calibration, fitting errors, and deblending issues.  We also examined
the SED of the source near GN26 to check the accuracy of the deblending.  The
$S_{70}/S_{160}$ ratio of the second source at $z\,{=}\,0.46$ is well fit by
the quiescently star-forming galaxy SED templates of \cite{dh02}, so the
deblending at 160$\,\mu$m seems reasonable. 

The multi-wavelength properties of the two SMGs which are detected at
70$\,\mu$m (Figure~1 and Table~1) are described in detail in Pope
et al.~(2006).  As expected, both sources (GN26 and GN13) are in the low
redshift tail of the submm redshift distribution.  Furthermore, we notice that
these two sources are among the faintest at 850$\,\mu$m in the submm sample
(both have $S_{850}\,{<}\,2.5\,$mJy) and therefore they are not typical of the
full submm sample presented in Pope et al.~(2006) or indeed other samples of
SMGs.  GN13 and GN26 have 70$\,\mu$m flux densities of 6.5 and $13.9\,$mJy,
respectively, while the full 70$\,\mu$m catalog has a median flux density of
$5\,$mJy.  About 80\% of the 70$\,\mu$m sources have spectroscopic redshifts,
and the median redshift of these sources is $z\,{=}\,0.46$ (Huynh et al.~in
preparation).  Thus the 70$\,\mu$m counterpart to GN13 is typical of the full
70$\,\mu$m sample, but GN26 is unusual in that it has a bright 70$\,\mu$m
counterpart, and is one of only seven 70$\,\mu$m sources in our sample currently confirmed
to be at $z\,{>}\,1$. 

\section{Stacking Analysis}

We performed a stacking analysis to derive an average 70$\,\mu$m flux density
for the SMG population in the GOODS-N field. 
To begin with, we stacked the {\sl Spitzer\/} data at the positions of all
SMGs, including sources with 70$\,\mu$m matches or with coincident flux.  For
each SMG a square image 132\arcsec\ on a side (approximately 7 MIPS beams) was
extracted.  We rotated each image by $90^\circ$ with respect to the previous
one before co-adding to remove any large-scale background effects.  The median
level of the individual extracted images was subtracted to remove any small
scale offsets and yield better background removal.  Flux densities were
determined using an aperture of 12\arcsec\ radius, and we applied an aperture
correction of 2.0, which was calculated empirically from bright sources in the
image. The aperture photometry was done after the background was subtracted from the stacked images, and the results were verified by making measurements with different size sky annuli.

To estimate the expected scatter, offset stacked images were generated by
randomly choosing a position in the 70$\,\mu$m image for each stacked source.
Five hundred such random stacks were generated ,and the uncertainty in the stacked flux
density is taken to be the standard deviation of these 500 measured values.

The average 70$\,\mu$m flux density ($\left\langle S_{70}\right\rangle$) for
all 30 SMGs is $2.00\pm0.48\,$mJy, and hence the stacked signal is detected at
over $4\sigma$.  The contribution from SMGs to the Extragalactic Background
Light (EBL) at 70$\,\mu$m can be estimated from this stacked signal by
multiplying by the appropriate source density.
Assuming the SMG integrated source count of
$2506\pm406\,{\rm deg}^{-2}$ for $S_{850}\,{>}\,2\,$mJy \citep{coppin2006}, we
estimate the contribution to the 70$\,\mu$m EBL from SMGs to be
$0.016\pm0.005\,{\rm MJy}\,{\rm sr}^{-1}$.  From an extrapolation of the source
counts \cite{frayer2006b} find that the total 70$\,\mu$m EBL is
$0.18\,{\rm MJy}\,{\rm sr}^{-1}$, so SMGs ($S_{850}\,{>}\,2\,$mJy) make up
about $9\pm3$\% of the 70$\,\mu$m EBL.

The 70$\,\mu$m stacked signal from all 30 SMGs is dominated by the 4 low
redshift sources with coincident 70$\,\mu$m flux density.  To determine the
average properties of the SMGs {\it not\/} detected at 70$\,\mu$m, which is
more representative of the general SMG population, we also stacked the 26
sources without coincident 70$\,\mu$m flux density.  This time we stacked the
residual 70$\,\mu$m image, which was obtained by removing all sources brighter
than 3$\sigma$ at 70$\,\mu$m.  Sources were removed by subtracting their
fitted point spread functions from the image.  The residual image was used to
obtain a better signal to noise ratio in the stacked image and in the measured
stacked flux density.  For these 26 sources we find
$\left\langle S_{70}\right\rangle\,{=}\,0.70\pm0.27\,$mJy, and so this
stacked signal has an SNR of about 3. 

To test whether the majority of the stacked 70$\,\mu$m flux density is from
the lower redshift SMGs, we also looked at low and high redshift sub-samples.
The low redshift sub-sample consists of 12 SMGs out of the 26 with $z\,{<}\,2$,
and the remaining 14 SMGs with $z\,{\geq}\,2$ make up the high redshift
sub-sample.  For redshift completeness we included IRAC photometric redshifts
from Pope et al.~(2006) for 8/26 SMGs, and these all have $z\,{\geq}\,1.8$
(and estimated accuracy of $\Delta z\,{\leq}\,0.4$).  The aperture flux density
in the central region of the high redshift stack is $0.22\pm0.44\,$mJy, while
for the low redshift stack it is more positive at $1.0\pm0.4\,$mJy.  This is
consistent with the idea that the majority of the flux density from the full
sample is coming from the lower redshift sources, although the measurements
are too noisy to make a definitive statement. 

Similarly, we stacked the 160$\,\mu$m image at all SMG positions, excluding the
one detected source, GN26.  There is no significant flux density in the stacked
image and the 3$\sigma$ upper limit is $13\,$mJy.  A stack of the 12 low
redshift SMGs gives a 3$\sigma$ upper limit at 160$\,\mu$m of $22\,$mJy.

To study the FIR properties of SMGs we could use the average flux densities of
the full sample of SMGs.  However, the stacked 70$\,\mu$m flux density from the
separate high- and low-$z$ sub-samples clearly show that most of the signal is
coming from $z\,{<}\,2$ sources (as expected from the K-correction).  We
therefore limit our analyses to the average properties of the low-$z$
sub-sample in the following sections.

\section{Infrared Colors}

The average IR colors, $S_{70}/S_{850}$ and $S_{70}/S_{24}$, are shown in
Figure~2 as a function of redshift.  GN26, at $z\,{\sim}\,1.5$ is consistent
with a more active Dale \& Helou (2002, hereafter DH02) model, with dust
intensity index $\gamma\,{=}\,1.5$, whereas GN13 has cooler colors,
corresponding to $\gamma\,{=}\,2.5$.  In the DH02 models $\gamma$ defines the
amount of dust as a function of heating intensity \citep{dale2001}:
\[ dM(U) \propto U^{-\gamma} dU \, ,\]
where $M(U)$ is the dust mass heated by a radiation field of intensity $U$.
These $\gamma$ values span the range expected for infrared luminous galaxies --
a low value around 1 implies a high contribution from photo-dissociation
regions in an actively star-forming galaxy, while a significantly higher
value represents the cirrus-dominated interstellar medium of a more quiescent
or cooler galaxy.

The infrared colors of the average low-$z$ SMG are consistent with
$\gamma\,{\simeq}\,2$--2.5 DH02 SEDs.  The average $S_{70}/S_{850}$ ratio of
SMGs from the stacking analysis indicates that the SMGs are relatively cool
galaxies for their high IR luminosities, which is consistent with
Pope et al.~(2006).  We also calculated upper limits to the $S_{70}/S_{850}$
ratio for SMGs {\it not\/} detected individually at 70$\,\mu$m (see Figure~2).
The lower redshift SMGs are clearly inconsistent with lower values of $\gamma$.
For the higher redshift sources we expect much lower $S_{70}/S_{850}$ ratios,
but with the current observational limits these sources can still be actively
star-forming.

For GN13, Figure~2 shows that the $S_{70}/S_{24}$ ratio is that of an actively
star-forming galaxy, with $\gamma\,{\simeq}\,1$--1.5, while the
$S_{70}/S_{850}$ ratio indicates it is a cooler galaxy.  This may be due to
broad silicate absorption falling into the 24$\,\mu$m band; the $S_{70}/S_{24}$
ratio can be strongly affected by PAH features and silicate absorption, which
are not fully accounted for in the models. 

We find that the colors of GN26 are consistent with DH02 models with
$\gamma\,{=}\,1$ and $z\,{\simeq}\,1$--2 (Figure~3).  The colors of GN13 place
it at $z\,{=}\,1$--2 if a $\gamma = 1.5$ model is assumed (Figure~3).  However,
GN13 is at $z\,{=}\,0.475$ and so this SMG has a warmer $S_{70}/S_{24}$ ratio
than that suggested by its $S_{70}/S_{850}$ ratio, as mentioned earlier.

The average colors of low-$z$ SMGs are also plotted in Figure~3.  We find that
the average IR colors are best represented with a DH02
model having $\gamma\,{\simeq}\,2$--2.5 at $z\,{=}\,1$--2.  This is consistent
with the median redshift of the low-$z$ SMG sub-sample, and suggests the
color-color plot can be used as a crude redshift indicator.

\section{Spectral Energy Distribution}

The FIR photometry at 70 and 160$\,\mu$m provides valuable data points for
constraining the FIR peak.  Combined with the 850$\,\mu$m observations, the
photometry spans both sides of the peak.  Previous estimates of the SED of SMGs
have relied on extrapolating the MIR or radio to fit the FIR peak -- at the
redshifts of SMGs, the MIR can be affected by complex emission and absorption
features, so this method may not be reliable.

We fit a variety of models to the data.  These include four DH02 models with
$\gamma$ = 1, 1.5, 2, and 2.5, as well as the Chary \& Elbaz (2001, hereafter
CE01) SEDs, which are templates derived from ISOCAM, {\sl IRAS\/} and SCUBA
data of nearby galaxies.  The CE01 models have luminosity dependent shapes, but
since the local luminosity--temperature relation found for nearby galaxies may
not hold for high-$z$ SMGs (e.g.~Pope et al.~2006), we fit CE01 models allowing
them to scale arbitrarily in luminosity.

For GN13 and for the low-$z$ SMG sub-sample, we constrain the fit with the 70
and 850$\,\mu$m observations only, since they are not detected at 160$\,\mu$m
(although the 160$\,\mu$m data provide a useful upper limit).  For GN26, the
70, 160 and 850$\,\mu$m observations are all used to constrain the fit.  We
summarize the fitting results in Table~2, while Figures~4 and 5 show the best
fit SEDs for GN13, GN26, and the average low-$z$ SMG.  Fitting CE01 SEDs
without allowing the luminosity to vary freely results in relatively poor fits,
so we exclude these from the Table.  This confirms the Pope et al.~(2006)
result that the local luminosity--temperature relationship does not hold for
SMGs.  For each best fit SED the total $L_{\rm IR}$ (between 8 and 1000 $\mu$m)
was calculated and given in Table~2.  Uncertainties in the luminosity were
derived by scaling the best fit SED until the minimum $\chi^2$ value exceeded
the 68\% ($\pm1\sigma$) confidence interval. 

The models provide good fits to the 70 and 850$\,\mu$m data for GN13 and for
the low-$z$ average SMG, but they do not typically fit the 24$\,\mu$m
(observed) data point, although it has been shown that the 24$\,\mu$m flux
density can often be fit with additional extinction (Pope et al.~2006).  The
models do not provide a similarly good fit in the FIR for GN26; the 70 and
850$\,\mu$m data points of GN26 are well fit, but the 160$\,\mu$m measurement
is underestimated by the models.  Hence the luminosity for GN26 is probably
higher than given by these models.  Nevertheless, our derived IR luminosity for
GN26 is 3 times greater than that estimated by Pope et al.~(2006) from fitting
SEDs to 24$\,\mu$m, 850$\,\mu$m and $1.4\,$GHz radio data.  It is possible that
there are further deblending issues at 160$\,\mu$m for GN26 (even although
we have already divided the total 160$\,\mu$m flux density between the two
70$\,\mu$m sources in the area).  This demonstrates the power of 160$\,\mu$m
photometry in constraining the total infrared luminosity of galaxies, but shows
that higher resolution is required to study such faint sources individually. 

The SMGs have high luminosities, but their FIR spectral shape is different from
local ULIRGs of the same luminosity.  We find that the average low-$z$ SMG has
a total IR luminosity of about $8.0\times10^{11}\,{\rm L}_{\odot}$.  This is a
factor ${\simeq}\,2$ less than the median SMG luminosity found by Pope et
al.~(2006) and Chapman et al.~(2005) for their SMGs with $z\,{<}\,2$.  The
reason that our calculated SMG luminosities are low compared to previous
results is because our best fit SEDs are cooler.  The average SMG is best fit
by a quiescent DH02 model with $\gamma\,{\simeq}\,2.5$, or with CE01 SED
templates of normal spiral galaxies scaled up by a factor ${\simeq}\,300$,
with a rest-frame peak at about $150\,\mu$m (i.e. $T\,{\simeq}\,20\,$K).  Local
ULIRGs of the same luminosity as the SMGs are therefore not the best spectral
templates for this sample.  

Several recent studies have relied on the MIR data (at 24$\,\mu$m in particular)
to derive luminosities and SED fits (e.g.~\citealp{perez2005},
\citealp{lefloch2005}).  The 24$\,\mu$m flux density was used in SED fitting of
SMGs by Pope et al.~(2006), who found that the typical SMG peaks at about
$100\,\mu$m (corresponding to about $29\,$K).  If the 24$\,\mu$m data point is
included in the fitting of GN13 along with the 70 and 850$\,\mu$m data, we find
the best fit DH02 SED peaks at shorter wavelength, corresponding to warmer dust
temperatures, and the total IR luminosity is decreased by a factor of about 2.
There is no significant difference in the CE01 fits to GN13 with and without
the 24$\,\mu$m data point.  For GN26 we find the 24$\,\mu$m data point makes no
significant difference to the best fit DH02 SED, while the best fit CE01 model
is slightly warmer and 3 times less luminous.  The fit to the average SMG
including the 24$\,\mu$m data point is warmer than that with only the longer
wavelength data, and the best fit DH02 and CE01 SEDs are about 2 times more
luminous.  At the median redshift of the average low-$z$ SMG,
$\left\langle z\right\rangle\,{=}\,1.4$, PAH and silicate features fall into
into the 24$\,\mu$m band, and our fit here is driven by these features
in the model SEDs.  The DH02 and CE01 SEDs certainly contain PAH features, but
it is not clear whether SMGs at this redshift have strong or weak PAH features,
if any.  Therefore we would argue that the fit to the 850 and 70$\,\mu$m flux
densities alone gives a more reliable result for the average SMG total
luminosity, at least for the moment, until we learn more about the MIR
spectra of SMGs.

\section{Dust Temperatures and Masses}

As a phenomenological alternative, we also adopt a modified blackbody SED
model to fit the temperature of these SMGs.  The SED is described by
$f_\nu \propto \nu^\beta B_\nu$, where $B_\nu(\nu, T)$ is the blackbody
function for dust of temperature $T$, and $\beta$ is the dust emissivity index.
The MIR is approximated as a power-law of the form $f_\nu\propto\nu^{-\alpha}$
and smoothly matches $\nu^\beta B_\nu$ at longer wavelengths \citep{blain2003}.
Although this simple phenomenological model cannot describe the full complex
dust properties of a galaxy, it can provide a good description of the general
behavior of the SED.  The range of parameters we consider is
$15\,{\rm K}\,{<}\,T\,{<}\,90\,{\rm K}$ and $1\,{<}\,\alpha\,{<}\,4$, which is
representative of galaxies ranging from normal spirals to AGN.  We set
$\beta\,{\equiv}\,1.5$ for our model fits, which is the value found for dust
in the galactic plane \citep{masi95}, and a typical value for well-studied
nearby galaxies \citep{dunne2000}.

We fit for $T$ and $\alpha$ using the 70$\,\mu$m and 850$\,\mu$m data points
for GN13 and the average low-$z$ SMG, but also include the 160$\,\mu$m
detection for GN26.  The results are summarized in Table~3.  When fitting only
two data points (and allowing the normalization to also be free) there is a
strong degeneracy between $\alpha$ and $T$ (which would be complete except for
the boundaries of the parameter ranges) -- so the fit parameters must be
interpreted with caution.   In the case of GN13, the full range of $\alpha$ is
allowed by the 70 and 850$\,\mu$m data, with low values of $\alpha$
corresponding to low values of $T$.  Because of the additional 160$\,\mu$m
data point, the parameters are better constrained for GN26, with
$T\,{\simeq}\,45\,$K and $\alpha\,{\simeq}\,3.5$ being preferred. 

For the average SMG, the 70 and 850$\,\mu$m flux densities alone cannot break
the $T$ and $\alpha$ degeneracy -- a very low temperature of $15\,$K is allowed
for $\alpha\,{\simeq}\,1.0$, while $T\,{\simeq}\,33\,$K for
$\alpha\,{\simeq}\,4.0$.  In a sample of 73 radio-detected SMGs the average
$S_{450}/S_{850}$ ratio is measured to be $5.0\pm2.3$ \citep{chapman2005},
while 15 SMGs from this same sample have been detected with SHARC-II (Kovacs et
al.~2006) and they have an average $S_{350}/S_{850}$ ratio of $4.0\pm1.3$.
We find that models with $\alpha\,{<}\,1.6$ are inconsistent with these
ratios.   This implies that the allowed models for the average SMG have
$21\,{\rm K}\,{<}\,T\,{<}\,33\,{\rm K}$ and $1.6\,{<}\,\alpha\,{<}\,4.0$,
where the low values of $T$ require low $\alpha$.  This shows that the low-$z$
SMGs have relatively cool dust temperatures. 

The best fit dust temperatures of 21--$33\,$K are consistent with those values
previously derived for SMGs (e.g.~Pope et al.~2006).  Chapman et al.~(2005)
and Kovacs et al.~(2006) suggested average temperatures close to the
upper end of our acceptable range.  This implies that the average low-$z$ SMG
in our sample has a relatively steep mid-IR SED, since our model fits with
$T\,{\simeq}\,30\,$K require $\alpha\,{\simeq}\,3$.  This suggests that the
SMGs are star-forming galaxies, because large $\alpha$ implies cool mid-IR
colors, which are inconsistent with AGN-dominated sources. 

Assuming that the submm light is thermal emission from dust which is optically
thin at $\lambda_{\rm rest}\,{\sim}\,200\,\mu$m, with a single dust
temperature $T$, the dust mass $M_{\rm d}$ is given by:
\begin{equation}
M_{\rm d} = \frac{S_{850} \: D^2_{\rm L}} {(1 + z) \:
 \kappa_{\rm d}(\nu_{\rm rest}) \: B_\nu(\nu_{\rm rest}, T)  }  \; ,
\end{equation}
(e.g.~\citealp{mcmahon1994}), where $D_{\rm L}$ is the cosmological luminosity
distance at redshift $z$ and the dust absorption coefficient $\kappa$ is
uncertain, even in the local Universe.  We take a $\kappa$ value of
$0.077\pm0.030\,{\rm m}^2{\rm kg}^{-1}$ \citep{hughes1993}, converting it to
rest-frame frequency $\nu_{\rm rest}$ with:
\begin{equation}
\kappa_{\rm d}(\nu_{\rm rest}) = 0.077
 \left( \frac{\nu_{\rm rest}}{350\,{\rm GHz}} \right)^\beta.
\end{equation}
Here we again assume that the dust emissivity index $\beta$ is fixed at 1.5.

The range of allowable dust masses are calculated from the range of
temperatures in Table~3.  The dust mass calculated from Equation~(2) is
1.0--1.6$\times10^8\,{\rm M}_\odot$ and 2.2--2.5$\times10^8\,{\rm M}_\odot$
for GN13 and GN26, respectively, while the dust mass found for the average 
low-$z$ SMG is 1.1--2.6$\times10^9\,{\rm M}_\odot$.  This does not take into
account the uncertainties in $\kappa$ and $S_{850}$; the dust mass uncertainty
is about 50\% when these are added in quadrature.  These dust masses are
consistent, within uncertainties, with the molecular gas mass derived from CO
observations (e.g.~\citealp{frayer1998}, \citealp{frayer99},
\citealp{neri2003}, \citealp{greve2005}), assuming a typical galactic gas mass
to dust mass ratio $M_{\rm g}/M_{\rm d}\,{\simeq}\,100$
(e.g.~\citealp{hildebrand1983}).
 
\section{FIR--Radio Correlation}

Our sample can be used to test the FIR--radio correlation in submillimeter
galaxies.  The FIR--radio correlation is often expressed as
(e.g.~\citealp{yun2001}):
\begin{equation}
q \equiv \log\left( \frac{{\rm FIR}}{ 3.75 \times 10^{12}\,{\rm W m}^{-2}}
 \right) -  \log\left( \frac{S_{1.4}}{{\rm W}\,{\rm m}^{-2}{\rm Hz}^{-1}}
 \right) \; ,
\end{equation}
where `FIR' here refers to the flux between 40 and 120 $\mu$m.  The observed
local value is $q\,{=}\,2.34\pm0.3$ (Yun, Reddy and Condon 2001).  We use the
best fit DH02 models to derive the conversion from $L_{40{-}120}$ to
$L_{8{-}1000}$, which is 2.0 and 1.6 for GN13 and GN26, respectively.
Based on their $L_{\rm IR}$ (Table \ref{lir_table}), we find $q$ parameters of $2.5^{+\,0.3}_{-\,0.1}$ and 
$2.4^{+\,0.1}_{-\,0.1}$ for GN13 and GN26, respectively.  These
values are consistent with the local value of $q$, suggesting that these two sources follow the local FIR--radio correlation.

\section{Contribution from Active Galactic Nuclei}

As mentioned in Section 8, the inferred high values of $\alpha$ suggest that
AGN do not dominate the bolometric luminosity in our sample of SMGs.  To
quantify the contribution of AGN to the infrared luminosity of SMGs, we adopt
the simple modified blackbody approach as described in Section~8.  For our AGN
model we use $\beta\,{=}\,1.5$, $T\,{=}\,90\,$K, and $\alpha\,{=}\,1.1$, as
found for the xFLS AGN population (Frayer et al.~2006).  We subtract this
AGN component from the observed 70 and 850$\,\mu$m flux densities of the
average SMG, and then repeat the fitting procedure with the DH02 and CE01
models, increasing the AGN component until the best fit $\chi^2$ value exceeds
the previous minimum by $1\sigma$.  This allows us to estimate the maximum AGN
contribution to the FIR luminosity, with the results shown in Figure~5.

It could be argued that the MIR waveband is the best discriminator of AGN, and
therefore we should be focussing on the 24$\,\mu$m data point
(e.g.~\citealp{sajina2005}).  However, we are here calculating the percentage
contribution of AGN to the $L_{\rm IR}$, which is dominated by the FIR peak,
and the contribution from AGN emission in the MIR is only a very small
proportion of the total IR luminosity. 

For the average low-$z$ SMG, an AGN which contributes up to 14\% of the IR
luminosity is allowed, using only the previous best fit CE01 SED model.  If all
CE01 models are used in the refitting, an AGN component of up to 23\%
contribution is allowed.  Together these fitting procedures imply that the
average (low-$z$ sub-sample) SMG is dominated by a starburst.  This is
consistent with X-Ray studies which find that AGN contribute on average 10\% to
the infrared luminosity of SMGs \citep{alexander2005}. 

For the 70$\,\mu$m detected SMGs, GN13 and GN26, we find that 32\% and 21\%,
respectively, of the total IR luminosity can be attributed to an AGN from the
same analysis applied to the best fit templates.  However, if the whole suite
of CE01 SEDs are used, a larger proportion of the IR luminosity can be
subtracted off the observed data points, with lower luminosity CE01 models
still fitting.  This shows that with the current data a low luminosity
starburst model from CE01 with an additional dominant AGN component is
indistinguishable from a high luminosity CE01 ULIRG model, and that further
photometry (or spectroscopy) is needed.

Another method to estimate the AGN contribution is to make use of the
24$\,\mu$m data.  Here, we consider the extreme case where all the 24$\,\mu$m
flux density is due to an AGN, and determine its contribution to the total IR
luminosity.  For GN13 and GN26 such an AGN would contribute 11\% and 5\%,
respectively, to the total IR luminosity, assuming the $T\,{=}\,90\,$K,
$\alpha\,{=}\,1.1$ AGN model used above, and the CE01 best fit SEDs.
For the average low-$z$ SMG, 21\% of the total IR luminosity could be
attributed to an AGN in this extreme case.  This again supports the hypothesis
that SMGs are dominated by star formation processes. 

\section{Concluding Remarks}

We have presented 70$\,\mu$m properties of submillimeter galaxies in the
GOODS-N field.  Out of 30 SMGs (with $S_{850}\,{>}\,1.7\,$mJy) in the overlap
region of this field, 2 are detected at relatively high significance in
ultra-deep (${\sim}\,0.6\,$ mJy rms) 70$\,\mu$m imaging.  Both of these
detected SMGs lie at relatively low redshift.  One SMG, GN26 at redshift
$z\,{=}\,1.2$, has infrared colors which indicate it is actively star forming.
The second SMG, GN13 ($z\,{=}\,0.47$) has infrared colors similar to normal
spirals, but with a much higher luminosity.  We confirm that these two SMGs
detected at 70$\,\mu$m follow the locally derived FIR-radio correlation. 

To determine the average properties of SMGs (most of which lie at $z\,{>}\,2$
and are not detected individually at 70$\,\mu$m), we performed a stacking
analysis and find that the average SMG has a 70$\,\mu$m flux density of
$0.70\pm0.27\,$mJy.  Most of the 70$\,\mu$m flux density is coming from the
lower redshift SMGs however.  We analysed the average properties of twelve
SMGs with $z\,{<}\,2$, which have a stacked 70$\,\mu$m flux density of
$1.0\pm0.4\,$mJy.  From a stack of all 30 SMGs in the ultra-deep 70$\,\mu$m
image, we find that the contribution of SMGs (with $S_{850}\,{>}\,2\,$mJy) to
the total Extragalactic Background Light at 70$\,\mu$m is $9\pm3$\%.

The average low-$z$ SMG ($\left\langle z\right\rangle\,{=}\,1.4$) has cool
infrared colors and an FIR SED that is best fit by a scaled up (about
300$\times$) normal spiral galaxy.  We also found that an AGN contributes 
less than 23\% of the total IR luminosity in SMGs. 

We find that the average low-$z$ SMG has an IR luminosity, $L_{8{-}1000}$,
of $8.0(\pm 2.2)\times10^{11}\,{\rm L}_{\odot}$.  The average low-$z$ SMG
therefore has a star formation rate of
$135\pm35\,{\rm M}_\odot\,{\rm yr}^{-1}$, using the relationship between SFR
and IR luminosity for starburst galaxies given by \cite{kennicutt1998}.
GN13 and GN26 have star-formation rates of $30\pm10$ and
$800\pm270\,{\rm M}_\odot\,{\rm yr}^{-1}$, respectively.  The median IR
luminosity of $z\,{<}\,2$ SMGs in Chapman et al.~(2005) is twice the value we
find for our average low-$z$ SMG, suggesting that theirs may have been an
overestimate, suffering from lack of FIR data.

The next generation submillimeter bolometer, SCUBA-2, will come online in 2007
\citep{holland2006}
and it is expected to yield a major improvement at 450$\,\mu$m,  where it
should reach a confusion limit of ${\sim}\,1\,$mJy ($5\sigma$).  This depth is
well-matched to that of deep {\sl Spitzer\/} 70$\,\mu$m imaging and should
provide much more overlap than the 850$\,\mu$m selected sources.  In addition,
SCUBA-2 will produce much larger samples of fainter 850$\,\mu$m sources, much
like the two 70$\,\mu$m detected submm sources in this study, and therefore
70$\,\mu$m data will be a valuable addition to understanding their SEDs. 
Also in the near future, the {\sl Herschel Space Observatory\/} will produce
confusion limited images across the wavelength range
75--500$\,\mu$m. {\sl Herschel\/} will therefore enable detailed study of the far-IR SEDs of large samples
of SMGs, which we have shown here to be feasible for a sub-set of SMGs using
{\sl Spitzer}.

\acknowledgements 

We thank the anonymous referee for helpful comments that improved this paper. 
MTH would like to thank Anna Sajina for helpful discussions. AP and DS
acknowledge support from the Natural Sciences and Engineering Research Council
of Canada and the Canadian Space Agency.  This work is based in part on
observations made with the {\sl Spitzer Space Telescope}, which is operated by
the Jet Propulsion Laboratory, California Institute of Technology under a
contract with NASA.  Support for this work was provided by NASA through an
award issued by JPL/Caltech.

\clearpage

\begin{table}
\caption{Summary of the properties of GN13, GN26 and the average low-$z$
SMG.}
\tiny
\begin{tabular}{lcccccccc}  \hline
galaxy &  RA & Dec & $S_{24}$ & $S_{70}$ & $S_{160}$ & $S_{850}$ & $S_{1.4}$ & $z$ \\ 
 &  J2000 & J2000 & ($\mu$Jy) & (mJy) & (mJy) & (mJy) & ($\mu$Jy) & \\
\hline  
GN13  & 12 36 49.72 & 62 13 12.8 & 371.0 $\pm$ 10.4 & 6.5 $\pm$ 1.3 & $<$43 & 1.9 $\pm$ 0.4 & 45.4 $\pm$ 5.4 & 0.475 \\
GN26 &  12 36 34.51 & 62 12 40.9 & 446.0 $\pm$ 5.1 &  13.9 $\pm$ 1.8 & 110 $\pm$ 27 & 2.2 $\pm$ 0.8 & 194.3 $\pm$ 10.4 & 1.219 \\
low-$z$ SMG (12 sources) & & & 258 $\pm$ 20 & 1.0 $\pm$ 0.4 & $<$22 & 4.0 $\pm$ 1.4 & $<$116 & 1.4 \\ \hline
\end{tabular}
\tablecomments {  The 70 and 160$\,\mu$m flux densities include the absolute calibration
error (of order 10\%).  The coordinates for GN13 and GN26 are IRAC positions (Pope et al.~2006).
The redshift given for the low-$z$ SMGs is the median for the sub-sample, and the
$S_{850}$ value given for the low-$z$ sub-sample is the weighted average. }
\label{smgprops}
\end{table}
 

\begin{table}
\centering
\caption {Summary of the best fit SEDs and the calculated total $L_{\rm IR}$.}
\begin{tabular}{lcccc}  \hline
galaxy & best fit DH02 & DH02 $L_{\rm IR}$ (${\rm L}_\odot$) & lum scaled CE01
 $L_{\rm IR}$   (${\rm L}_\odot$) \\ \hline  
GN13 & $\gamma = 2.5$ & $(2.5\pm0.3)\times10^{11}$ & ($1.8\pm0.2)\times10^{11}$ \\
GN26 & $\gamma = 1.5$ & $(4.5\pm0.7)\times10^{12}$ & ($5.0\pm0.7)\times10^{12}$ \\
low-$z$ SMG (12 sources) & $\gamma = 2.5$ & ($9.0\pm2.5)\times10^{11}$ & $(8.0\pm2.2)\times10^{11}$  \\ \hline
\end{tabular}
\label{lir_table}
\end{table}


\begin{table}
\centering
\caption{Allowable model parameters for a simple modified blackbody SED. }
\begin{tabular}{lccccc}  \hline
galaxy & $T$(K) & $\alpha$ \\ \hline
 GN13  & 34 -- 50 & 1.0 -- 4.0   \\
 GN26 &  44 -- 48  & 3.0 -- 4.0  \\
 low-$z$ SMG (12 sources) &  21 -- 33 & 1.6 -- 4.0  \\ \hline
\end{tabular}
\tablecomments{ The range in
$T$ and $\alpha$ indicates the models which fit within the $1\sigma$ limit of
the 70, 160 and 850$\,\mu$m observations.  The dust emissivity parameter,
$\beta$, is assumed to be 1.5.}
\end{table}

\clearpage

\begin{figure}
\centering
\includegraphics[width=5cm]{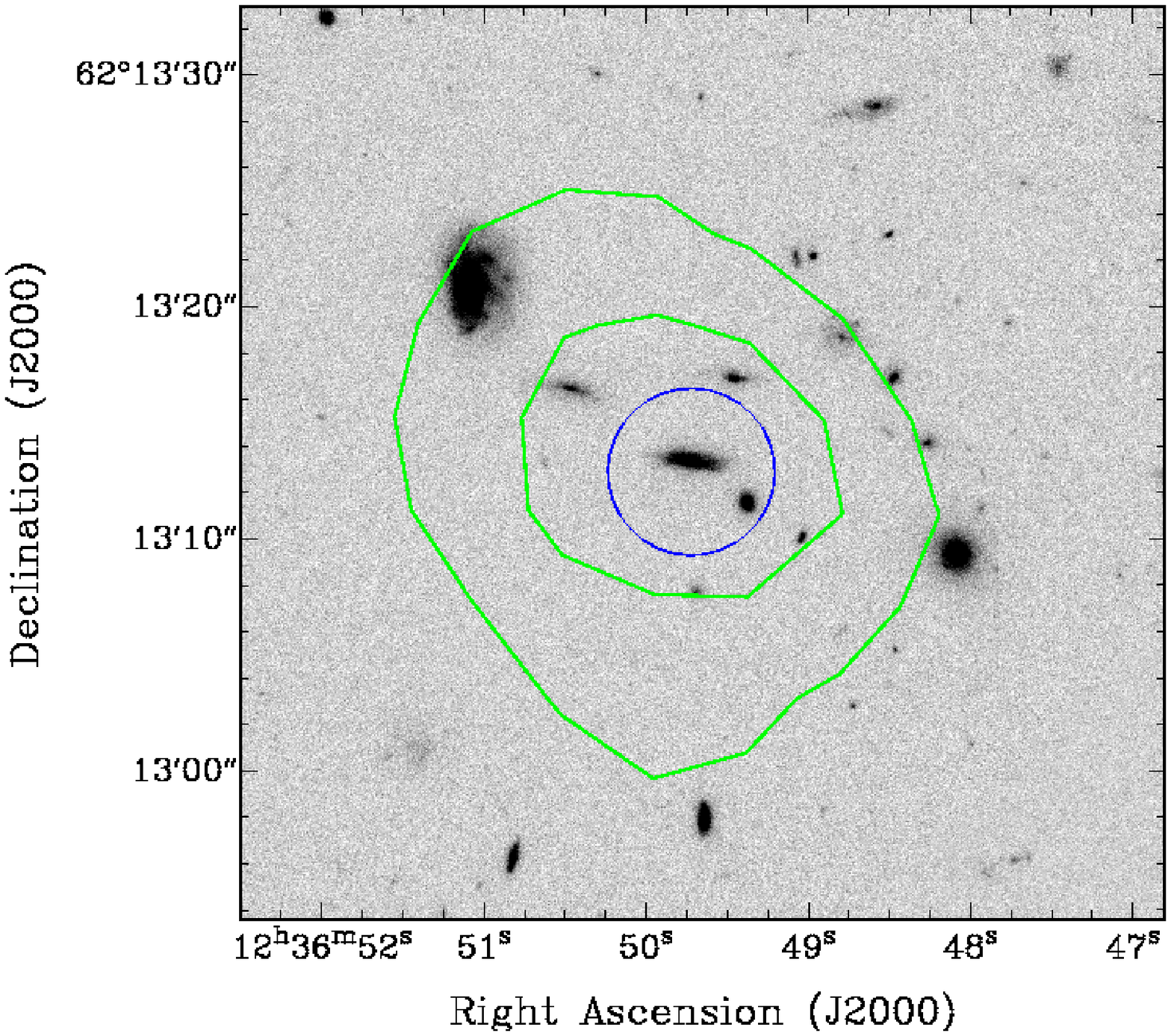}
\includegraphics[width=5.3cm]{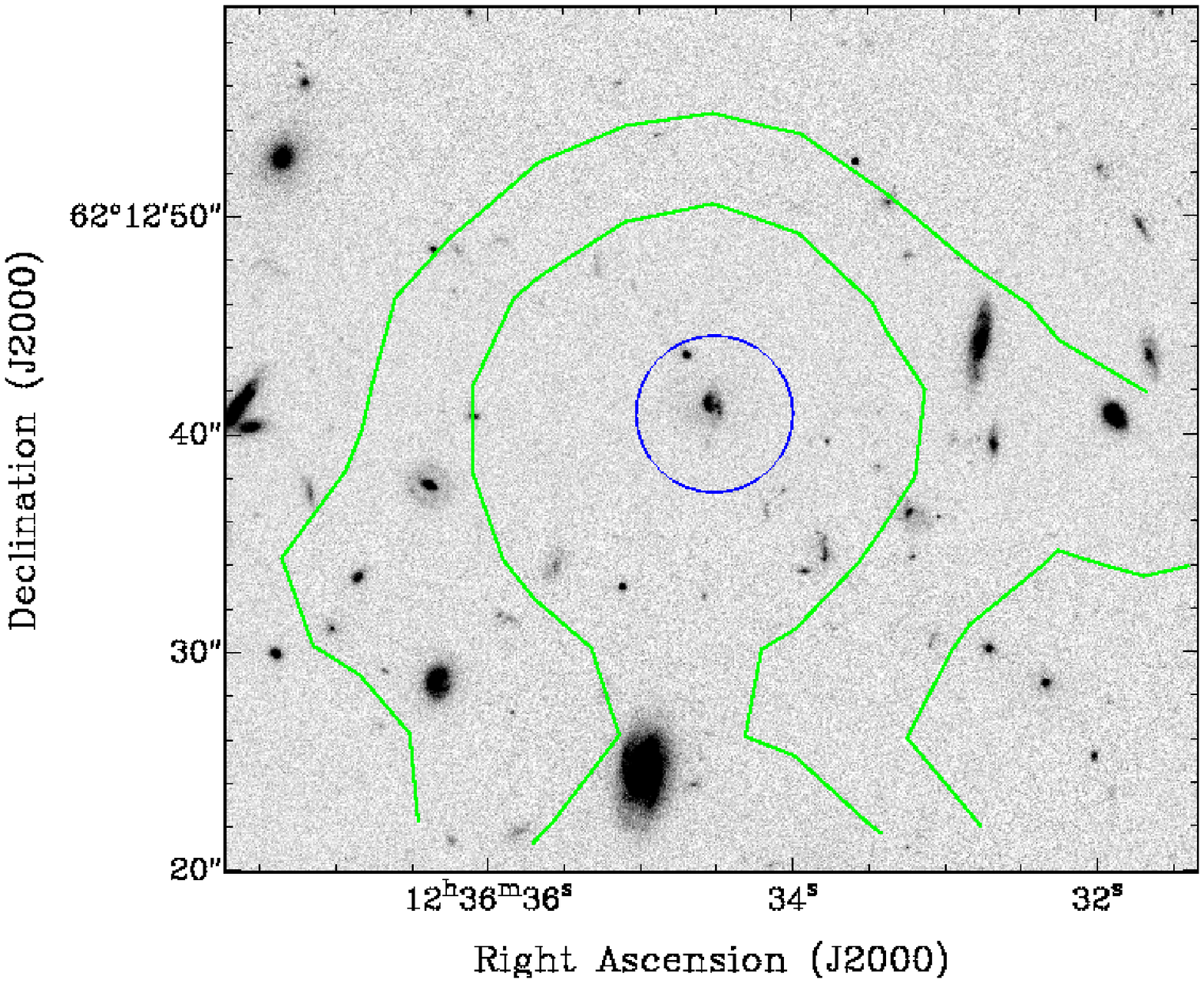}
\caption{Greyscale {\sl HST\/} ACS F606W images of GN13 (left panel) and GN26
(right panel).  The contours show 70$\,\mu$m data, plotted at $3\sigma$ and
$6\sigma$ levels.  The circle indicates the Pope et al.~(2006) IRAC counterpart
to the SCUBA source.  The images are $40^{\prime\prime}\times40^{\prime\prime}$ in size.}
\end{figure}

\clearpage

\begin{figure}
\centering
\includegraphics[width=10cm]{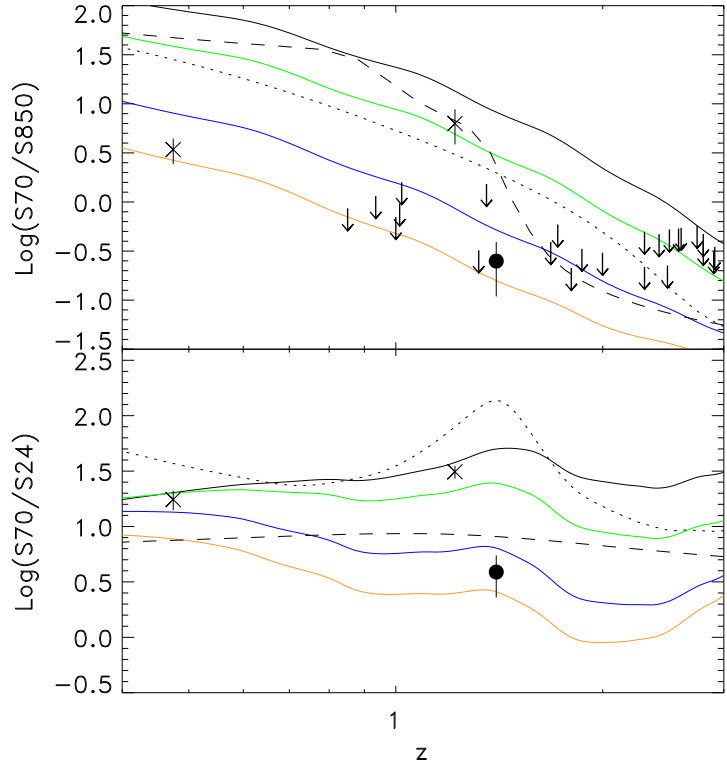}
\caption{Top Panel: $S_{70}/S_{850}$ colors of the DH02 models as a function of
redshift.  From top to bottom, the solid lines are $\gamma = 1$, 1.5, 2, and
2.5.  The dotted line marks an Arp220 SED, while the dashed line is Mrk231.
The crosses mark the SMGs GN13 and GN26, which are detected at 70$\,\mu$m.
Other SMGs are shown as $3\sigma$ upper limits.  The filled circles denote the
average ratios from the stacked analysis of the low-$z$ sub-sample at
$\left\langle z\right\rangle\,{=}\,1.4$.  Bottom Panel: $S_{70}/S_{24}$ colors
of the DH02 models. Lines and symbols are as for the top panel.}
\end{figure}

\clearpage

\begin{figure}
\centering
\includegraphics[width=10cm]{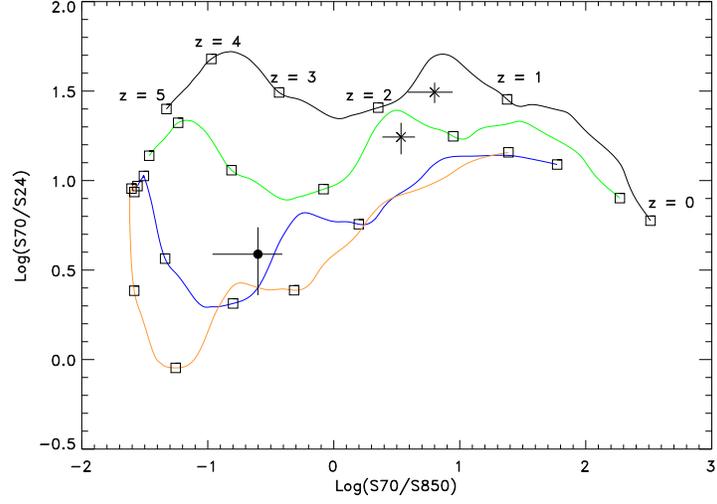}
\caption{$S_{70}/S_{24}$ versus $S_{70}/S_{850}$ colors compared with the DH02
models.  For each model the squares mark redshift 0, 1, 2, 3, 4, and 5, from
right to left.  The models from top to bottom are for $\gamma = 1$, 1.5, 2, and
2.5.  The crosses mark the SMGs GN13 and GN26, which are detected at
70$\,\mu$m.  The filled circle marks the average for the low-$z$ SMG
sub-sample.}
\end{figure}

\clearpage

\begin{figure}
\centering
\includegraphics[width=13cm]{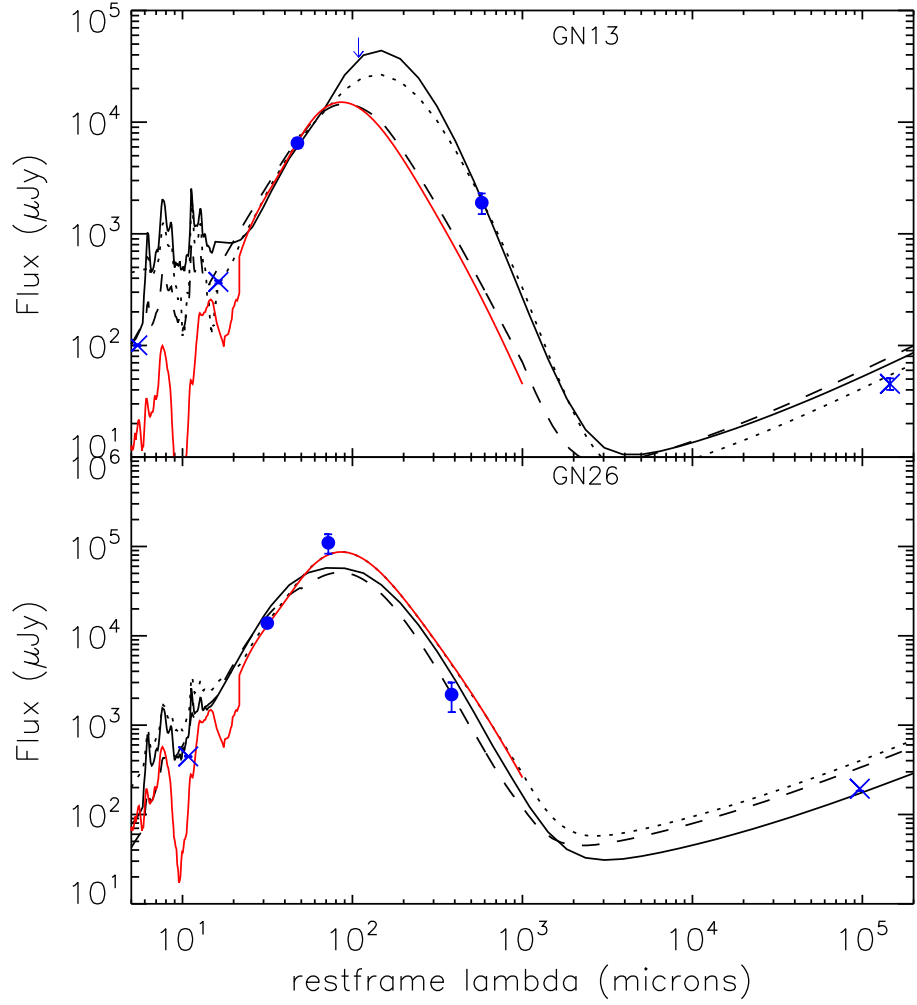}
\caption{SEDs of GN13 and GN26.  The observed
{\sl Spitzer\/} and SCUBA data are plotted at rest-frame wavelengths.
The best fit DH02 model is plotted as a solid black line, while the dashed
curve is the best fit CE01 luminosity dependent SED and the dotted line is the
best fit CE01 SED with luminosity allowed to vary.  The best fit Arp220 SED
is also shown as the red solid line.
The fits were constrained using only by the 70 and 850$\,\mu$m data for
GN13, and with the addition of the 160$\,\mu$m data point for GN26.}
\end{figure}

\clearpage

\begin{figure}
\centering
\includegraphics[width=10cm]{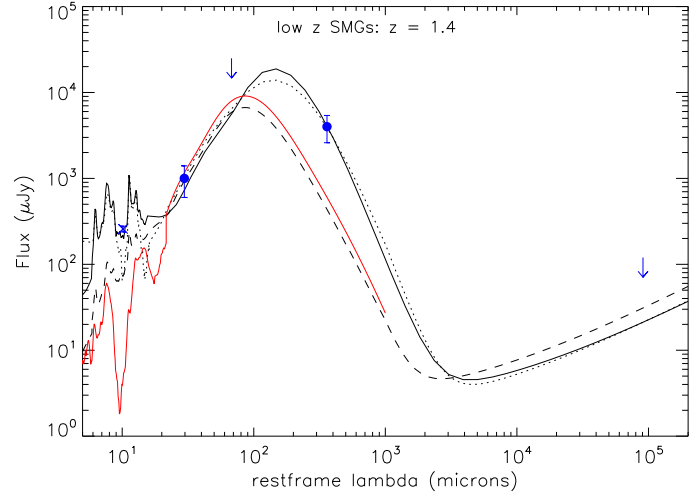}
\caption{The SED of the average low-$z$ SMG.  The observed {\sl Spitzer\/} and
SCUBA data are plotted at rest-frame wavelengths and the curves are:
best fit DH02 model (solid black line); best fit CE01 luminosity-dependent SED
(dashed line); best fit CE01 SED with luminosity allowed to vary (dotted line);
and best fit Arp220 SED (red solid line).  The fits were constrained using
the 70 and 850$\,\mu$m data only, as described in the text.}
\end{figure}

\clearpage

\begin{figure}
\centering
\includegraphics[width=10cm]{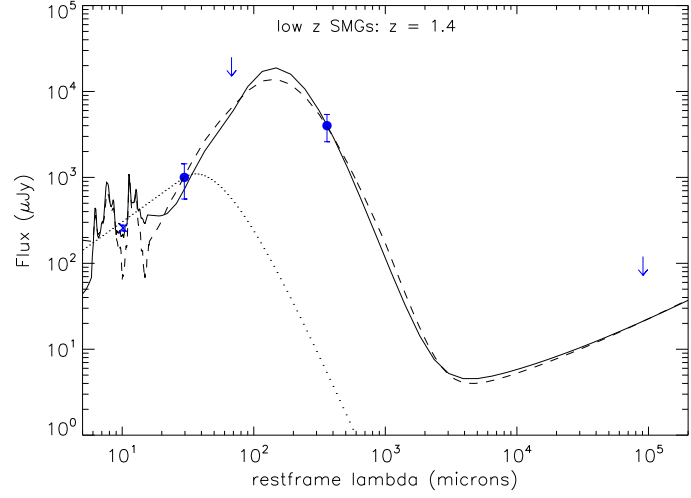}
\caption{The best fit SEDs for the low-$z$ SMG sub-sample.  The solid line is
the DH02 model and the dashed  line is the best fit CE01 model (allowing the
luminosity to float freely).  The dotted line shows the AGN contribution
allowed before all the CE01 models fail at the $1\sigma$ level.  The symbols
are as for Figure~5. }
\end{figure}


\begin{thebibliography}{}

\bibitem[\protect\citeauthoryear{{Alexander} et~al.}{{Alexander}
  et~al.}{2005}]{alexander2005}
{Alexander}, D.~M., {Bauer}, F.~E., {Chapman}, S.~C., {Smail}, I., {Blain},
  A.~W., {Brandt}, W.~N.,  \& {Ivison}, R.~J. 2005, \apj, 632, 736

\bibitem[\protect\citeauthoryear{{Barger}, {Cowie}, \& {Richards}}{{Barger}
  et~al.}{2000}]{barger2000}
{Barger}, A.~J., {Cowie}, L.~L.,  \& {Richards}, E.~A. 2000, \aj, 119, 2092

\bibitem[\protect\citeauthoryear{{Barger} et~al.}{{Barger}
  et~al.}{1998}]{barger1998}
{Barger}, A.~J., {Cowie}, L.~L., {Sanders}, D.~B., {Fulton}, E., {Taniguchi},
  Y., {Sato}, Y., {Kawara}, K.,  \& {Okuda}, H. 1998, \nat, 394, 248

\bibitem[\protect\citeauthoryear{{Blain}, {Barnard}, \& {Chapman}}{{Blain}
  et~al.}{2003}]{blain2003}
{Blain}, A.~W., {Barnard}, V.~E.,  \& {Chapman}, S.~C. 2003, \mnras, 338, 733

\bibitem[\protect\citeauthoryear{{Blain} et~al.}{{Blain}
  et~al.}{2002}]{blain2002}
{Blain}, A.~W., {Smail}, I., {Ivison}, R.~J., {Kneib}, J.-P.,  \& {Frayer},
  D.~T. 2002, \physrep, 369, 111

\bibitem[\protect\citeauthoryear{{Borys} et~al.}{{Borys}
  et~al.}{2002}]{borys2002}
{Borys}, C., {Chapman}, S.~C., {Halpern}, M.,  \& {Scott}, D. 2002, \mnras,
  330, L63

\bibitem[\protect\citeauthoryear{{Chapman} et~al.}{{Chapman}
  et~al.}{2005}]{chapman2005}
{Chapman}, S.~C., {Blain}, A.~W., {Smail}, I.,  \& {Ivison}, R.~J. 2005, \apj,
  622, 772

\bibitem[\protect\citeauthoryear{{Chapman} et~al.}{{Chapman}
  et~al.}{2002}]{chapman2002}
{Chapman}, S.~C., {Lewis}, G.~F., {Scott}, D., {Borys}, C.,  \& {Richards}, E.
  2002, \apj, 570, 557

\bibitem[\protect\citeauthoryear{{Chapman} et~al.}{{Chapman}
  et~al.}{2004}]{chapman2004}
{Chapman}, S.~C., {Smail}, I., {Blain}, A.~W.,  \& {Ivison}, R.~J. 2004, \apj,
  614, 671

\bibitem[\protect\citeauthoryear{{Coppin} et~al.}{{Coppin}
  et~al.}{2006}]{coppin2006}
{Coppin}, K., et~al. 2006, ArXiv Astrophysics e-prints, astro-ph/0609039

\bibitem[\protect\citeauthoryear{{Coppin} et~al.}{{Coppin}
  et~al.}{2005}]{coppin2005}
{Coppin}, K., {Halpern}, M., {Scott}, D., {Borys}, C.,  \& {Chapman}, S. 2005,
  \mnras, 357, 1022

\bibitem[\protect\citeauthoryear{{Dale} \& {Helou}}{{Dale} \&
  {Helou}}{2002}]{dh02}
{Dale}, D.~A.,  \& {Helou}, G. 2002, \apj, 576, 159

\bibitem[\protect\citeauthoryear{{Dale} et~al.}{{Dale} et~al.}{2001}]{dale2001}
{Dale}, D.~A., {Helou}, G., {Contursi}, A., {Silbermann}, N.~A.,  \&
  {Kolhatkar}, S. 2001, \apj, 549, 215

\bibitem[\protect\citeauthoryear{{Dole} et~al.}{{Dole} et~al.}{2004}]{dole2004}
{Dole}, H., et~al. 2004, \apjs, 154, 87

\bibitem[\protect\citeauthoryear{{Downes} et~al.}{{Downes}
  et~al.}{1986}]{downes1986}
{Downes}, A.~J.~B., {Peacock}, J.~A., {Savage}, A.,  \& {Carrie}, D.~R. 1986,
  \mnras, 218, 31

\bibitem[\protect\citeauthoryear{{Dunne} et~al.}{{Dunne}
  et~al.}{2000}]{dunne2000}
{Dunne}, L., {Eales}, S., {Edmunds}, M., {Ivison}, R., {Alexander}, P.,  \&
  {Clements}, D.~L. 2000, \mnras, 315, 115

\bibitem[\protect\citeauthoryear{{Frayer} et~al.}{{Frayer}
  et~al.}{2006a}]{frayer2006}
{Frayer}, D.~T., et~al. 2006a, \aj, 131, 250

\bibitem[\protect\citeauthoryear{{Frayer} et~al.}{{Frayer}
  et~al.}{2006b}]{frayer2006b}
{Frayer}, D.~T., et~al. 2006b, \apjl, 647, L9

\bibitem[\protect\citeauthoryear{{Frayer} et~al.}{{Frayer}
  et~al.}{1999}]{frayer99}
{Frayer}, D.~T., et~al. 1999, \apjl, 514, L13

\bibitem[\protect\citeauthoryear{{Frayer} et~al.}{{Frayer}
  et~al.}{1998}]{frayer1998}
{Frayer}, D.~T., {Ivison}, R.~J., {Scoville}, N.~Z., {Yun}, M., {Evans}, A.~S.,
  {Smail}, I., {Blain}, A.~W.,  \& {Kneib}, J.-P. 1998, \apjl, 506, L7

\bibitem[\protect\citeauthoryear{{Gordon} et~al.}{{Gordon}
  et~al.}{2005}]{gordon2005}
{Gordon}, K.~D., et~al. 2005, \pasp, 117, 503

\bibitem[\protect\citeauthoryear{{Greve} et~al.}{{Greve}
  et~al.}{2005}]{greve2005}
{Greve}, T.~R., et~al. 2005, \mnras, 359, 1165

\bibitem[\protect\citeauthoryear{{Greve} et~al.}{{Greve}
  et~al.}{2004}]{greve2004}
{Greve}, T.~R., {Ivison}, R.~J., {Bertoldi}, F., {Stevens}, J.~A., {Dunlop},
  J.~S., {Lutz}, D.,  \& {Carilli}, C.~L. 2004, \mnras, 354, 779

\bibitem[\protect\citeauthoryear{{Hildebrand}}{{Hildebrand}}{1983}]{hildebrand%
1983}
{Hildebrand}, R.~H. 1983, \qjras, 24, 267

\bibitem[\protect\citeauthoryear{{Holland} et~al.}{{Holland}
  et~al.}{2006}]{holland2006}
{Holland}, W., et~al. 2006, in Millimeter and Submillimeter Detectors and
  Instrumentation for Astronomy III. Edited by Zmuidzinas, Jonas; Holland,
  Wayne S.; Withington, Stafford; Duncan, William D.. Proceedings of the SPIE,
  Volume 6275, pp. (2006).

\bibitem[\protect\citeauthoryear{{Holland} et~al.}{{Holland}
  et~al.}{1999}]{holland1999}
{Holland}, W.~S., et~al. 1999, \mnras, 303, 659

\bibitem[\protect\citeauthoryear{{Hughes} et~al.}{{Hughes}
  et~al.}{1993}]{hughes1993}
{Hughes}, D.~H., {Robson}, E.~I., {Dunlop}, J.~S.,  \& {Gear}, W.~K. 1993,
  \mnras, 263, 607

\bibitem[\protect\citeauthoryear{{Hughes} et~al.}{{Hughes}
  et~al.}{1998}]{hughes1998}
{Hughes}, D.~H., et~al. 1998, \nat, 394, 241

\bibitem[\protect\citeauthoryear{{Ivison} et~al.}{{Ivison}
  et~al.}{2000}]{ivison2000}
{Ivison}, R.~J., {Smail}, I., {Barger}, A.~J., {Kneib}, J.-P., {Blain}, A.~W.,
  {Owen}, F.~N., {Kerr}, T.~H.,  \& {Cowie}, L.~L. 2000, \mnras, 315, 209

\bibitem[\protect\citeauthoryear{{Kennicutt}}{{Kennicutt}}{1998}]{kennicutt199%
8}
{Kennicutt}, R.~C. 1998, \araa, 36, 189

\bibitem[\protect\citeauthoryear{{Klaas} et~al.}{{Klaas}
  et~al.}{1997}]{klaas1997}
{Klaas}, U., {Haas}, M., {Heinrichsen}, I.,  \& {Schulz}, B. 1997, \aap, 325,
  L21

\bibitem[\protect\citeauthoryear{{Kreysa} et~al.}{{Kreysa}
  et~al.}{1998}]{kreysa1998}
{Kreysa}, E., et~al. 1998, in Proc. SPIE Vol. 3357, p. 319-325, Advanced
  Technology MMW, Radio, and Terahertz Telescopes, Thomas G. Phillips; Ed., ed.
  T.~G. {Phillips}, 319

\bibitem[\protect\citeauthoryear{{Lagache}, {Puget}, \& {Dole}}{{Lagache}
  et~al.}{2005}]{lagache2005}
{Lagache}, G., {Puget}, J.-L.,  \& {Dole}, H. 2005, \araa, 43, 727

\bibitem[\protect\citeauthoryear{{Le Floc'h} et~al.}{{Le Floc'h}
  et~al.}{2005}]{lefloch2005}
{Le Floc'h}, E., et~al. 2005, \apj, 632, 169

\bibitem[\protect\citeauthoryear{{Masi} et~al.}{{Masi} et~al.}{1995}]{masi95}
{Masi}, S., et~al. 1995, \apj, 452, 253

\bibitem[\protect\citeauthoryear{{McMahon} et~al.}{{McMahon}
  et~al.}{1994}]{mcmahon1994}
{McMahon}, R.~G., {Omont}, A., {Bergeron}, J., {Kreysa}, E.,  \& {Haslam},
  C.~G.~T. 1994, \mnras, 267, L9

\bibitem[\protect\citeauthoryear{{Neri} et~al.}{{Neri} et~al.}{2003}]{neri2003}
{Neri}, R., et~al. 2003, \apjl, 597, L113

\bibitem[\protect\citeauthoryear{{P{\'e}rez-Gonz{\'a}lez}
  et~al.}{{P{\'e}rez-Gonz{\'a}lez} et~al.}{2005}]{perez2005}
{P{\'e}rez-Gonz{\'a}lez}, P.~G., et~al. 2005, \apj, 630, 82

\bibitem[\protect\citeauthoryear{{Pope} et~al.}{{Pope} et~al.}{2005}]{pope2005}
{Pope}, A., {Borys}, C., {Scott}, D., {Conselice}, C., {Dickinson}, M.,  \&
  {Mobasher}, B. 2005, \mnras, 358, 149

\bibitem[\protect\citeauthoryear{{Pope} et~al.}{{Pope} et~al.}{2006}]{pope2006}
{Pope}, A., et~al. 2006, \mnras, 370, 1185

\bibitem[\protect\citeauthoryear{{Sajina}, {Lacy}, \& {Scott}}{{Sajina}
  et~al.}{2005}]{sajina2005}
{Sajina}, A., {Lacy}, M.,  \& {Scott}, D. 2005, \apj, 621, 256

\bibitem[\protect\citeauthoryear{{Sajina} et~al.}{{Sajina}
  et~al.}{2006}]{sajina2006}
{Sajina}, A., {Scott}, D., {Dennefeld}, M., {Dole}, H., {Lacy}, M.,  \&
  {Lagache}, G. 2006, \mnras, 518

\bibitem[\protect\citeauthoryear{{Smail}, {Ivison}, \& {Blain}}{{Smail}
  et~al.}{1997}]{smail1997}
{Smail}, I., {Ivison}, R.~J.,  \& {Blain}, A.~W. 1997, \apjl, 490, L5

\bibitem[\protect\citeauthoryear{{Webb} et~al.}{{Webb} et~al.}{2003}]{webb2003}
{Webb}, T.~M., et~al. 2003, \apj, 587, 41

\bibitem[\protect\citeauthoryear{{Yun}, {Reddy}, \& {Condon}}{{Yun}
  et~al.}{2001}]{yun2001}
{Yun}, M.~S., {Reddy}, N.~A.,  \& {Condon}, J.~J. 2001, \apj, 554, 803

\end{thebibliography}
\end{document}